\documentclass[conference]{IEEEtran}
\usepackage{float}
\usepackage{graphicx}
\usepackage[misc,geometry]{ifsym}

\begin{document}
\title{Towards a Modelling Framework for Self-Sovereign Identity Systems}
%
%

\author{\IEEEauthorblockN{
Iain Barclay\Letter\IEEEauthorrefmark{1},
Maria Freytsis\IEEEauthorrefmark{2},
Sherri Bucher\IEEEauthorrefmark{2}\IEEEauthorrefmark{3},
Swapna Radha\IEEEauthorrefmark{4},\\
Alun Preece\IEEEauthorrefmark{1}, 
Ian Taylor\IEEEauthorrefmark{1}\IEEEauthorrefmark{4}}
\IEEEauthorblockA{\IEEEauthorrefmark{1}School of Computer Science and Informatics, Cardiff University, UK\\ Email: BarclayIS@cardiff.ac.uk}
\IEEEauthorblockA{\IEEEauthorrefmark{2}NeoInnovate Collaborative Consortium, Indiana University School of Medicine, Indianapolis, IN, USA}
\IEEEauthorblockA{\IEEEauthorrefmark{3}Eck Institute for Global Health, University of Notre Dame, Notre Dame, IN, USA}
\IEEEauthorblockA{\IEEEauthorrefmark{4}Center for Research Computing, University of Notre Dame, Notre Dame, IN, USA}
}

\maketitle          

\begin{abstract}
Self-sovereign Identity promises to give users control of their own data, and has the potential to foster advancements in terms of personal data privacy. Self-sovereign concepts can also be applied to other entities, such as datasets and devices. Systems adopting this paradigm will be decentralised, with messages passing between multiple actors, both human and representing other entities, in order to issue and request credentials necessary to meet individual and collective goals. Such systems are complex, and build upon social and technical interactions and behaviours. Modelling self-sovereign identity systems seeks to provide stakeholders and software architects with tools to enable them to communicate effectively, and lead to effective and well-regarded system designs and implementations.  This paper draws upon research from Actor-based Modelling to guide a way forward in modelling self-sovereign systems, and reports early success in utilising the iStar 2.0 framework to provide a representation of a birth registration case study.

\end{abstract}
\section{Introducing Self-Sovereign Identity}

The term Self-Sovereign Identity (SSI)\cite{allen2016path} is used to describe the ability of an individual to take ownership of their personal data and to control who has access to that data, without the need for a centralised infrastructure, or any control or authorization being required by any third party. SSI has been the subject of research and ambition for several years, but has reached an inflection point in interest from industry and the research community as a result of the availability of distributed ledger and blockchain-based technologies, combined with an increased focus on individual's data privacy as they interact with web-based and social networking services\cite{tobin2016inevitable}. The COVID-19 pandemic, and a desire by global stakeholders and partners to develop creative technological solutions by which to address the various epidemiological, public health, and economic
impacts of regional and local outbreaks, has kindled additional interest in the self-sovereign concept and
its nascent implementations.

SSI is decentralised, and is built upon well-established cryptographic techniques whereby a user holds a private and shares a public key\cite{preneel1994cryptographic}. The private key is used to sign documents, whilst the public key can be used by anybody with access to it to verify that the document has indeed been signed, and has not been tampered with. SSI uses a system built on decentralised identifiers (DIDs) to identify parties involved, with the DIDs resolving to documents which explain, via machine-readable documents, how to locate the public key needed to validate claims made about that DID, in the same way as web addresses resolve to provide web pages. The SSI research community has developed data models and protocols\cite{sporny2019} that provide mechanisms for any party identified by a DID to issue cryptographically verifiable sets of credentials to any subject entity, also identified by a DID. In this way, a party which believes something to be true about another party can declare this in a standardised way using a JSON-LD\cite{lanthaler2012using} formatted document, and sign this attestation using asymmetric cryptography techniques, based on the DIDs used being able to be resolvable to validate the assertions made. This cryptographically signed document is known as a Verifiable Credential (VC), and will be held by the subject of the credential, or in the case of a child, or dataset or IoT asset, by an authorised Holder. 

\begin{figure}
\centering
\includegraphics[width=0.45\textwidth]{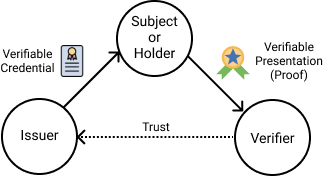}
\caption{The Triangle of Trust between parties} \label{figtrust}
\end{figure}

At a later date, when the holder seeks to enter into a transaction, a service provider may request proof of status or entitlements. The Verifiable Credential document provides a means for this proof to be provided, as the holder of the credential can generate a Verifiable Presentation containing assertions from the VC document. By processing the Verifiable Presentation document, the Verifier can use the accessible public keys to check that i) the presented proof pertains to the subject it is being presented on behalf of, ii) the presented proof contains assertions signed by the original Issuer, and finally iii) that the presented documents have not been tampered with. As such, triangles of trust\cite{davie2019trust} can be leveraged to enable parties to issue, hold and verify credentials without reliance on any central authority, as illustrated in Figure\ref{figtrust}.

\section{Modelling Self-Sovereign Identity Systems}
To date, the focus of effort in the SSI community has been on personal identity and data privacy for individuals\cite{wang2020self}, however the underlying computer science techniques can be applied to any type of entity, including digital assets such as datasets\cite{barclay2020}, and physical devices partaking in the Internet of Things (IoT)\cite{bartolomeu2019self}. Systems based on the paradigm of the self-sovereignty of human participants, data resources and IoT devices are inherently decentralised, with attributes held at the edges of the ecosystem, with access to them granted on request. These systems mix interactions with people and machines to achieve the desired goals, and are an embodiment of socio-technical systems. Access to data is granted at the discretion of the holders of the data, in response to requests from parties which require access to complete a transaction or meet a goal. From the perspective of a verifying party checking credentials of a holder, a request to present a proof of credentials could be required for two reasons:

\begin{itemize}
\item \emph{Qualities} -- ``Does the party that I’m interacting with have the qualities that I need in order for me to use it to complete my task?'' -- for example, does a software product that I'm evaluating have a suitable level of certification?

\item \emph {Entitlements} -- ``Does the party that I’m interacting with hold the entitlements they need allow them to access resources that I control?'' -- e.g., do they have the correct levels of authority to enter the building?
\end{itemize}

In both cases, the Verifying party requires proof of the credentials being held by the other party. In order to gain this proof, the Verifier makes a request to the Holder, for a presentation of the credentials. The Holder can make a decision on whether they are comfortable to comply with the request and provide a presentation. On receipt of the presentation of the credentials, the Verifier can check validity, and make their own judgement on the suitability of the credentials being presented. The consideration of this judgement will depend on the value of the interaction, but will typically take into account the identity and status of the Issuer of the credentials that are being presented, such that credentials will need to have been issued by credible or trusted parties in order to be acceptable. Furthermore, additional proof may be required to show that the credential has been issued to the party presenting it\cite{hardman2019using}.

As such, SSI systems can be seen to involve multiple actors, with their own sets of both personal and organisational goals, and with policies which need to be verified and conditionally enacted. The system operates through interactive communication and messages between the actors, in requesting to access resources or services, and in being asked to issue, or to provide proofs of credentials. 

Current instantiations of SSI systems are immature, with standards going through the W3C recommendation process\cite{sporny2019} and commercial projects largely at the pilot or proof of concept stage. Implementations of such systems are complex, requiring infrastructure for cryptographic key generation and management, secure credential storage, and mechanisms for secure message passing between the parties, both between agents in the cloud and mobile wallet applications held and interacted with by end-users. As such, there is currently little practical experience and shared knowledge of how to design and build these systems available for researchers and engineers to draw upon, and implementation is time consuming and reliant on skilled developer resources. 

In spite of practical implementation difficulties, the mechanisms of operation of SSI systems are clear, and defined by published protocols. It seems desirable, therefore, for SSI researchers to use models to further their understanding of the mechanics of the systems they seek to design, such that effective implementations can be delivered once the underlying infrastructure matures. It is hoped that providing clear and concise documentation and models of these complex systems will promote effective communication between domain experts, software architects and developers. The prospect of converting models into simulations of SSI systems, operating in agent-based environments, such as Hash.ai\footnote{https://hash.ai}, is an interesting prospect. This would enable interactions between the actors to be visualised by stakeholders, and systems to analysed and verified and validated to ensure that policies are being correctly expressed and implemented. 

In seeking techniques for modelling SSI systems, literature from the field of Actor Based Modelling (ABM) has offered a promising starting point. In particular, the i* framework developed in the PhD thesis of Yu\cite{yu2011modeling} in the mid-nineties, and developed further in recent years towards the iStar 2.0\cite{dalpiaz2016istar} modelling language, offers many parallel constructs to those explored in SSI systems, and provides a lens through which techniques for modelling of SSI systems can begin to be investigated. 

\begin{figure}[ht]
\centering
\includegraphics[width=0.45\textwidth]{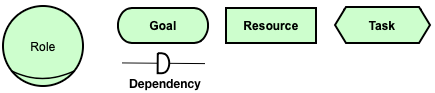}
\caption{Key iStar 2.0 Symbols} \label{fig0}
\end{figure}

In particular, iStar 2.0 provides metaphors to define the Actors in a system, and to specify actor's goals as well as any tasks they can perform, along with resources they have access to - as detailed in Figure \ref{fig0}. Furthermore, iStar 2.0 allows for dependencies between actors to be specified, allowing expression of the need for actors in the system to have reliance on each other to perform their goal. Such dependencies take the form of goals that actors need other actors to enact, tasks they are dependent on other actors to perform or resources they need other actors to provide them with. In particular, the means of expressing a dependency on actors in the system providing other actors with resources promises a good alignment with parties in an SSI system issuing credentials to other parties, and requesting proofs of credentials in order to satisfy the goals of one party or another.

iStar 2.0 offers two main perspectives on an ecosystem being modelled -- a Strategic Dependency View which shows dependencies between the actors, and a Strategic Rationale View which provides a means to look inside each of the actors to see internal goals, tasks and resources. A further, Hybrid View, provides a combination of the other two views, with internals of some actors hidden and other's exposed, as appropriate to the part of the ecosystem under analysis.

\section{Case Study: A Birth Registration Process}
As means to gain exposure to the iStar 2.0 framework and modelling process and to begin to gauge its applicability to describing SSI systems, an illustrative case study is explored. This case study comes from the NeoLinkID project, part of the NeoInnovate Collaborative at Indiana University, and considers a birth registration process. The scenario described below is being studied as a hypothetical venue for the replacement of paper documents with personally held digital documentation in an urban area of Kenya. A simplified version of the Kenyan birth registration process can be considered to have the following steps:

\begin{enumerate}
\item The midwife present at the birth populates a Birth Notification Document
(BND). As part of this process, the mother of the child presents their iden-
tity documents. The mother's name, and identity number, along with other
details of the birth are entered into the BND, which is given to the baby's
mother for safe-keeping. A copy for the registration form is sent to the District Registrar’s Office. 

\item  At a later date, the mother visits the District Registrar’s Office. The Registrar checks the BND and the mother's identity documentation against the copy received in the office, and if all is in order, starts the process for issuing a birth certificate to the mother for her child.
\end{enumerate}

Digital birth registration is a sought-after use case in the areas of global health and development, humanitarian relief, and digital identity. The ability to accurately model this use case can allow stakeholders to gain a better understanding of dependencies and facilitate a shared language for evaluating digital birth registration systems that use verifiable credentials and public key cryptography.

\subsection{Strategic Dependency View}
\begin{figure}[ht]
\centering
\includegraphics[width=0.5\textwidth]{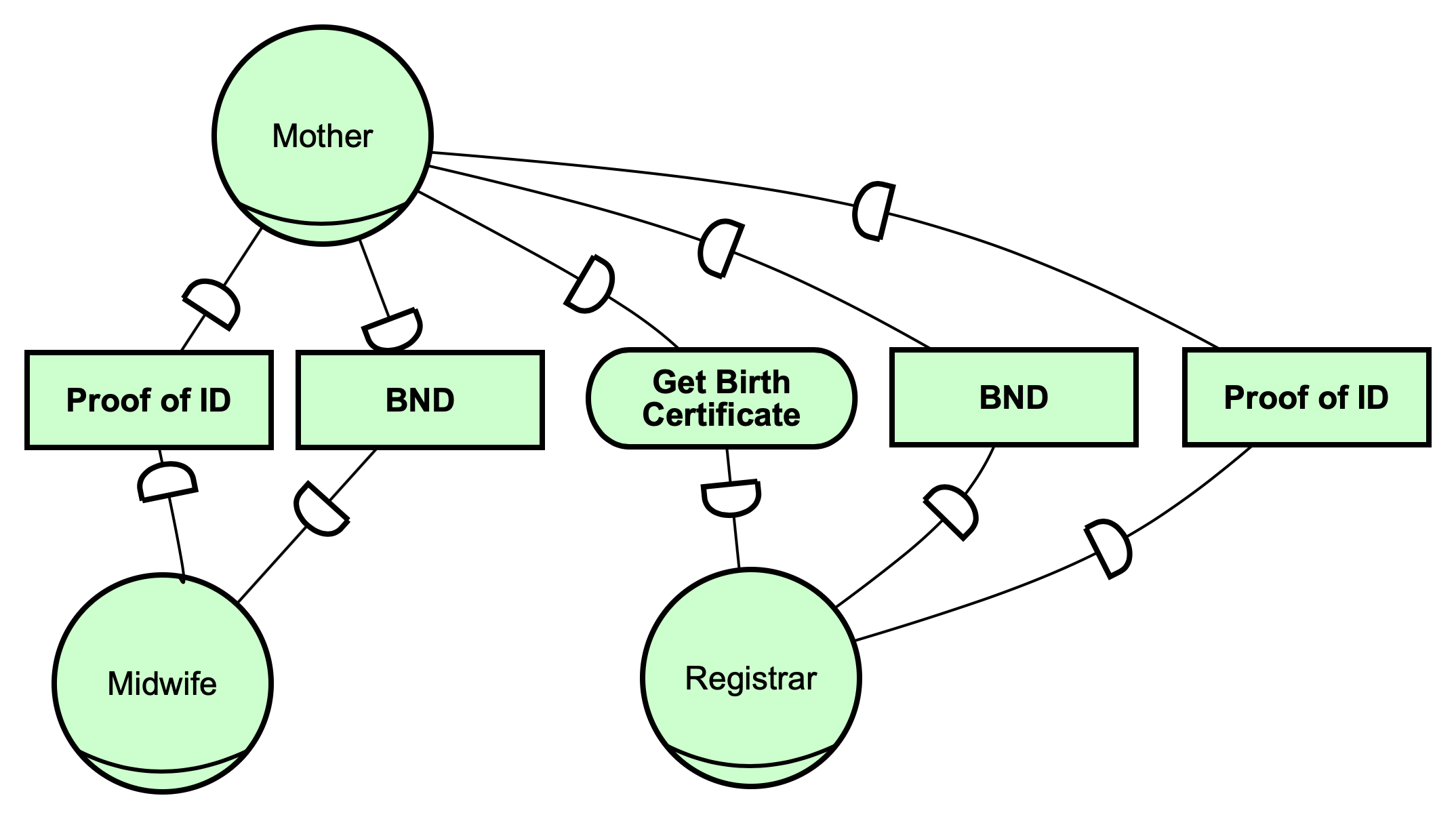}
\caption{A Strategic Dependency view of birth registration.} \label{fig1}
\end{figure}

Entities from the case study description have been rendered as iStar 2.0 components, using the piStar\cite{pimentel2018pistar} web-based tool as a means to capture information. In the first instance, the iStar 2.0 Strategic Dependency (SD) View (Figure \ref{fig1}) illustrates the dependencies that exist between the different Actors in the case study. In an SSI system, the Actors are the parties that would request and issue credentials, and execute and respond to credential proof requests. The SD Graph highlights the Mother's goal of Get Birth Certificate, which is reliant on provision on the BND and the Mother's ID to the Registrar. In turn, the BND needs to have been provided to the Mother by the Midwife, so that it can be presented later.

The SD View for the simplified birth registration process provides a clear mapping to the elements in a self-sovereign design for the system, where individual actors control their own data and credentials, with proofs of credentials being requested and provided as appropriate.

The mapping between actors in the SD graph and the roles played in SSI systems is that the Midwife is a Verifier of the Mother's ID, as well as being the Issuer of the BND credential. The direction of the Dependency markers on the diagram illustrate which Resources need to be presented to each party, or in self-sovereign terms, where a proof of credential would need to be presented. It is not clear, however, whether a Resource needs to be issued to an Actor, or if it needs to be presented by the Actor. Figure \ref{fig1} shows that the Mother needs the BND to be issued to her by the Midwife, whereas the Midwife needs the Mother to present proof of already held credentials (i.e. proof of ID credentials issued by another agency). Both are shown in the same way in the iStar 2.0 SD diagram.

\subsection{Strategic Rationale View}
\begin{figure*}[ht]
\centering
\includegraphics[width=0.8\textwidth]{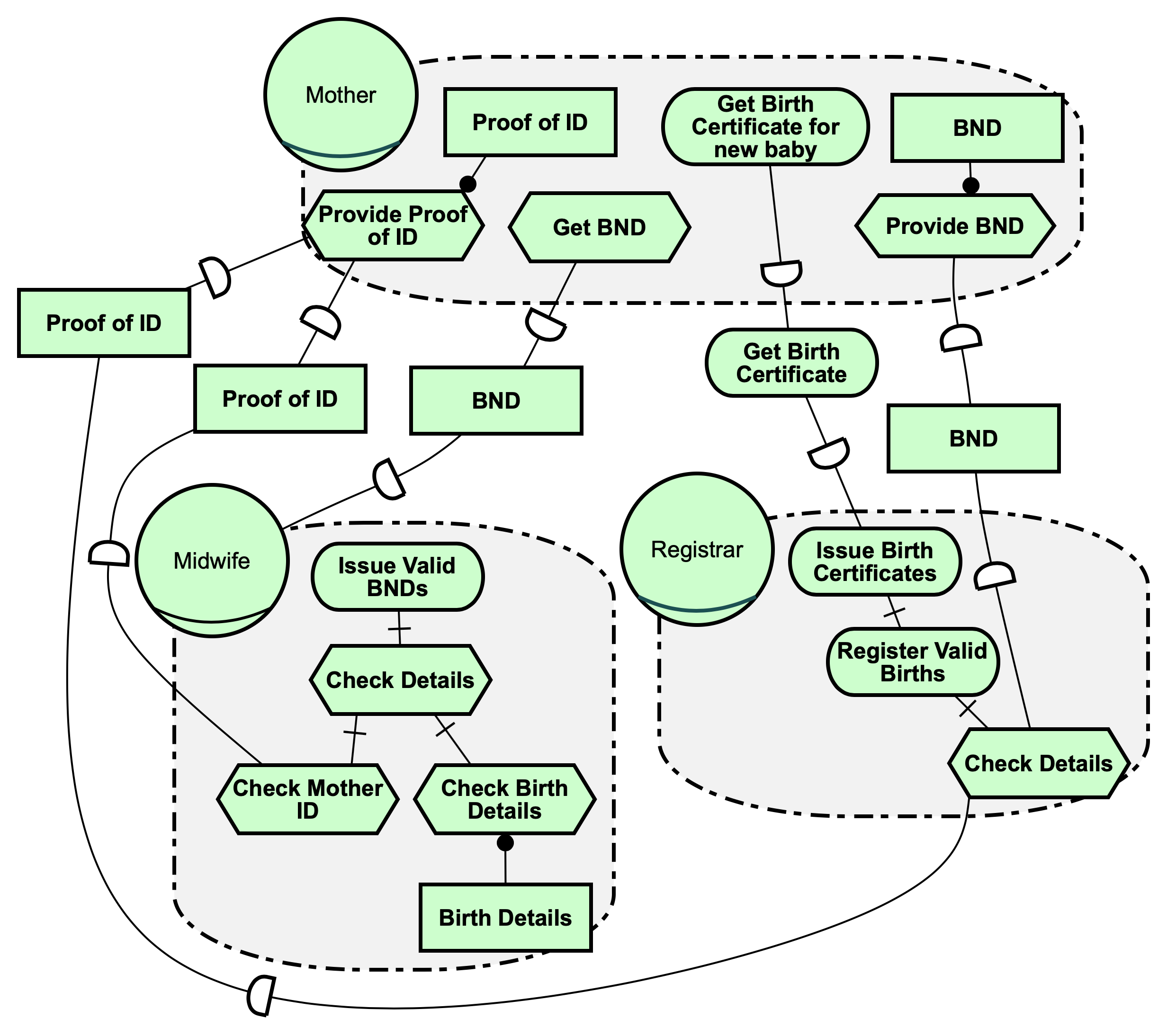}
\caption{A Strategic Rationale view of birth registration.} \label{fig2}
\end{figure*}
The iStar 2.0 Strategic Rationale (SR) View (Figure \ref{fig2}) looks at the goals and tasks inside each Actor, and considers what needs to be achieved in order to meet their goals and any requests made of them by other actors in the ecosystem. The defining goal of the case study can be seen inside the Mother role, which is to ``Get Birth Certificate for new baby''. This goal depends on the Registrar's goal of ``Issue Birth Cerificates'', which in turn requires the Registrar to check both the Mother's ID and the BND - the BND in turn, needs to be provided to the Mother by the Midwife, whose own goal is to ``Issue Valid BNDs''. Looking inside of each Actor beings to give insight into the logic of the processes and the requirements and constraints in the system. The use of terminology such as ``Check'' inside an Actor is a good indicator that the Actor is a Verifier of a presented credential, whereas ''Provide'' can be used to give clarity that the Actor will be presenting proof of a credential they hold, and ``Issue'' that they are a party responsible for issuing a credential to another Actor. As such, the mapping of the each actor's roles in the SSI system can be well understood from the Strategic Rationale View. Additionally, the SR View provides scope for the inclusion of policies and governance frameworks, which are a crucial non-technical piece of the self-sovereign identity model. Such frameworks describe the trust models for SSI systems, determining relationships between credential issuing parties, for example, such that Verifiers are able to judge which credentials should be accepted and which not.

\section{Conclusion and Future Work}
Modelling the birth registration use case with iStar 2.0 and framing a visual representation with the piStar tool has provided an insightful graphical depiction of the Actors in the case study, their goals, and the resources (as Credentials and Proofs) that need to pass between them. Considering the effectiveness of the iStar 2.0 model of the birth registration use case, it is noted that the Strategic Dependency view doesn't clearly provide indication whether a party is responsible for issuing a credential or for presenting a proof of a credential that they already hold. At this early investigative stage, this would appear to be a shortcoming in the modelling that would have an impact in describing how an implementation of the system would be developed.

It is possible that modifications to the primitives used either for the representation of Actors, or for the interface between the Resource dependency and the Actor to show that the Actor needs to issue a credential for this Resource would address this, or that further experience modelling with the iStar 2.0 language or related frameworks might lead to a solution and recommendation based on the existing language. When looking deeper into the modelling, via the Strategic Rationale view, the use of language such as ``Issue'', ``Provide'' and ``Check'' is able to clearly and simply describe whether the action is one of credential issuance, proof presentation or verification. 

In order to evaluate further, the iStar 2.0 framework will be applied to additional use cases, including those specific to the COVID-19 pandemic, to enable broader insight of its applicability for modelling SSI systems. Modelling of the same systems using other goal-oriented frameworks will also take place, such that comparisons can be drawn in order to determine the most effective approach for future analysis. 
A longer-term and important goal of research into modelling SSI systems is to identify or develop an approach to converting models into software agents, such that they can be rendered and used in a simulation environment, with the intent of allowing system behaviours to be experienced and refined, and interactions to be tested prior to development and deployment. As such, it will be extremely desirable if the tools used to model the systems are able to output code that is able to seed simulations with the appropriate actor and message interfaces. This, in turn, might kindle additional progress toward practical SSI applications.

\bigskip \noindent\textbf{Acknowledgements.}
The authors would like to record their thanks to Fabiano Dalpiaz for providing helpful feedback on early versions of this paper and suggestions towards future research goals and directions.

\newpage

%
%
%
\bibliographystyle{splncs04}
\bibliography{modelling}

\end{document}